\documentclass{article}
\usepackage[utf8]{inputenc}

\title{Shaping of Magnetic Field Coils in Fusion Reactors using Bayesian Optimisation}
\usepackage{graphicx}
\usepackage{caption}
\usepackage{xcolor}
\usepackage{float}
\usepackage{url}
\usepackage[final]{gpsmdms_2022}

\author{
  Timothy Nunn \\
  UK Atomic Energy Authority\\
  Oxford, OX14 3EB \\
  \texttt{timothy.nunn@ukaea.uk} \\
  \and
  Vignesh Gopakumar \\
  UK Atomic Energy Authority \\
  Oxford, OX14 3EB \\
  \texttt{vignesh.gopakumar@ukaea.uk} \\
  \and
  Sebastien Kahn \\
  UK Atomic Energy Authority \\
  Oxford, OX14 3EB \\
  \texttt{sebastien.kahn@ukaea.uk} 
}

\begin{document}

\maketitle

\begin{abstract}
Nuclear fusion using magnetic confinement holds promise as a viable method for sustainable energy. However, most fusion devices have been experimental and as we move towards energy reactors, we are entering into a new paradigm of engineering. Curating a design for a fusion reactor is a high-dimensional multi-output optimisation process. Through this work we demonstrate a proof-of-concept of an AI-driven strategy to help explore the design search space and identify optimum parameters. By utilising a Multi-Output Bayesian Optimisation scheme, our strategy is capable of identifying the Pareto front associated with the optimisation of the toroidal field coil shape of a tokamak. The optimisation helps to identify design parameters that would minimise the costs incurred while maximising the plasma stability by way of minimising magnetic ripples. 
\end{abstract}

\section{Introduction}
A tokamak is plasma magnetic confinement system laid out in a toroidal structure~\cite{wesson2004tokamaks}. The goal is to confine the plasma at sufficiently high temperature to generate fusion reactions, attempting to generate electricity. To efficiently design a tokamak, a wide variety of physics and engineering challenges must be resolved within a limited amount of space. Hence, the process governing such design often involves various simulation codes and multiple stakeholders, often drawn out for months, while only evaluating a limited section of the design space \cite{FEDERICI2018729}. The tokamak design code Bluemira, resulting from the merge between the BLUEPRINT~\cite{COLEMAN201926} and the MIRA~\cite{Franza_2022} codes, integrates various aspects of a fusion power plant design in the same framework, significantly reducing the design optimization time, allowing a wider design exploration. A major component of a tokamak is its toroidal field (TF) coils, responsible for the generation of the magnetic pressure confining the plasma. We invite the reader to view Appendix \ref{app:appendix_tokamak} for a schematic diagram of a Tokamak including the typical location of the TF coils. This work proposes a first use of an AI-based exploration and identification of optimum fusion reactor component design. We deploy a Multi-Objective Bayesian Optimisation algorithm (MOBO) to optimise  the shape of the TF coils. For a further reading about TF coil optimization, we invite the reader to read the following thesis~\cite{Coleman_phd}. The MOBO is deployed over Bluemira and benchmarked against conventional optimizers to demonstrate the efficacy of our approach, laying the foundation to extend this proof-of-concept to actively search more complex design spaces.

\section{Problem} 
\label{sec:problem}

The TF coil shape influences various aspects of the tokamak design, from the distribution of the electromechanical forces (structural design) to the size of the reactor (cost) and the toroidal field variations (ripple) at the plasma boundaries driving its performance~\cite{ripple_effects, Narl_Davidson_1976}. The problem considers optimising across the parameterisation of the 3D shape of the TF coils conductors (number, thickness, height, radii) as outlined in appendix \ref{app:appendix_params} and illustrated in figure~\ref{fig:app:TF_coil_shape_parametrization}. The considered coil conductors have a rounded rectangular shape, while its cross-section is taken to be rectangular for simplicity. The optimisation is performed with the objective metrics summarised in table~\ref{tab:opt_metrics}, that outline the volume of the magnets and its induced ripple effects. The magnetic field is calculated using an analytical 3D magnetostatic formulation, allowing one to quickly and accurately calculate the magnetic field at an arbitrary point in space, using a series of current filaments to approximate the conductor geometry of the TF coils. For a given point in the poloidal plane, the ripple is defined as the relative difference between the maximum magnetic field (on the TF coil leg) and the minimum magnetic field (between two TF coil legs) as described in~\cite[p. 46]{Coleman_phd}. The coil size is simply defined as the perimeter length multiplied by the conductor cross-section. The plasma shape is fixed for this experiment as defined by the geometric parameterisation described in~\cite[p. 310-312]{Johner_helios}. The number of points at which the ripple is evaluated along the flux surface (ripple points) is fixed, ensuring that ripple evaluations occur at the same point on the plasma for each design exploration.

\section{Optimisation}

The optimisation pipeline was built using the BoTorch \cite{botorch} framework integrated with GpyTorch\cite{gpytorch}. The implementation of the design search space exploration using a Bayesian framework was done using the Ax API \cite{Bakshy2018AEAD} built on top of BoTorch and GpyTorch. 

\subsection{Setup}
In order to optimise the shape of TF coils using a Bayesian strategy, the problem is framed as a multi-objective optimisation scenario. The optimisation requires searching an 8-dimensional parameter hyperspace to find ideal points on the Pareto front that would minimise the magnetic ripple effects as well as the volumetric size of the TF coils as shown in Table \ref{tab:opt_metrics}. The parameters to be optimised for and their domain ranges can be found in Appendix \ref{app:appendix_params}. Within the Bayesian optimisation process we have chosen to avoid adding additional constraints and have instead instilled the geometric constraints in the parameter bounds. 

\begin{table}[h]
    \begin{center}
        \begin{tabular}{|c|c|}
            \hline
            Metric & Description \\
            \hline\hline
            Ripple & Maximum value of magnetic ripple encountered while evaluating the \emph{ripple} function \\
            \hline
            Size & Volumetric sum of the TF Coils. \\
            \hline
        \end{tabular}
        \caption{The metric name and description for each of the metrics being minimised.}
         \label{tab:opt_metrics}
    \end{center}
\end{table}

\subsection{Implementation}

The optimisation is split into two portions: initial quasi-random sampling within the domain followed by multi-objective Bayesian optimisation (MOBO) across the design space. The initial sampling was done using the Sobol \cite{SOBOL196786} quasi-random sampling strategy. The MOBO loop will run for a prescribed number of iterations. Each iteration, a series of Gaussian Process (GP) surrogates are trained on all the currently observed data \cite{rasmussen2005gaussian}. We construct an individual GP for each objective using Exact GPs, as the ripple function from Bluemira is treated to be as ground truth without any noise. Each GP is defined using a Gaussian likelihood and a prior with constant mean and a Matern Kernel \cite{rasmussen2005gaussian} using a length scale prior. GPs are fitted by minimising the sum marginal log likelihood across them. Inputs to the GPs are each mapped independently onto the unit hypercube such that any design point $X \in [0,1]^n$ where this problem has $n$ inputs. Outputs $Y$ are standardised to have a mean 0 and variance 1. Within the Bayesian framework, newer design points are sampled using the expected hypervolume improvement (EHVI) \cite{ehvi} acquisition function across the fitted GPs, in line with the mentioned objectives.

\section{Results}
\subsection{Comparison of designs}
Because we are minimising both \emph{size} and \emph{ripple} of the design, there is no objective best design point. Instead we get a Pareto frontier of design points. The feasible design points are ordered by \emph{size} and, generally, the design with the smallest coil is selected. 

\begin{table}[h!]
    \centering
    \begin{tabular}{|c|c|c|c|c|}
         \hline
          Expt \# & Sobol Samples & Bayesian Samples & Size & Ripple \\
        \hline\hline
            1 & 64 & 20 & 0.46 & 0.018\\
        \hline
            2 & 128 & 40 & 0.091 & 4.5e-4\\
        \hline
    \end{tabular}
    \caption{Comparison of the optimal design solution(s) achieved on different experiments utilising the Sobol + Bayesian optimisation method.}
    \label{tab:sobol_baysopt_comp}
\end{table}

The Gaussian process is able to understand the general profile of the two metrics relatively quickly. Some GP fits associated with each objective can be seen in Appendix \ref{appendix: gp_fit}. As seen in Table \ref{tab:sobol_baysopt_comp}, experiment 2 finds better quality points, by several orders of magnitude, than experiment 1. Generally, as more samples of the \emph{ripple} function are evaluated, points closer to the "true" Pareto front are found.

\begin{figure}
    \centering
    \includegraphics[width=0.8\textwidth]{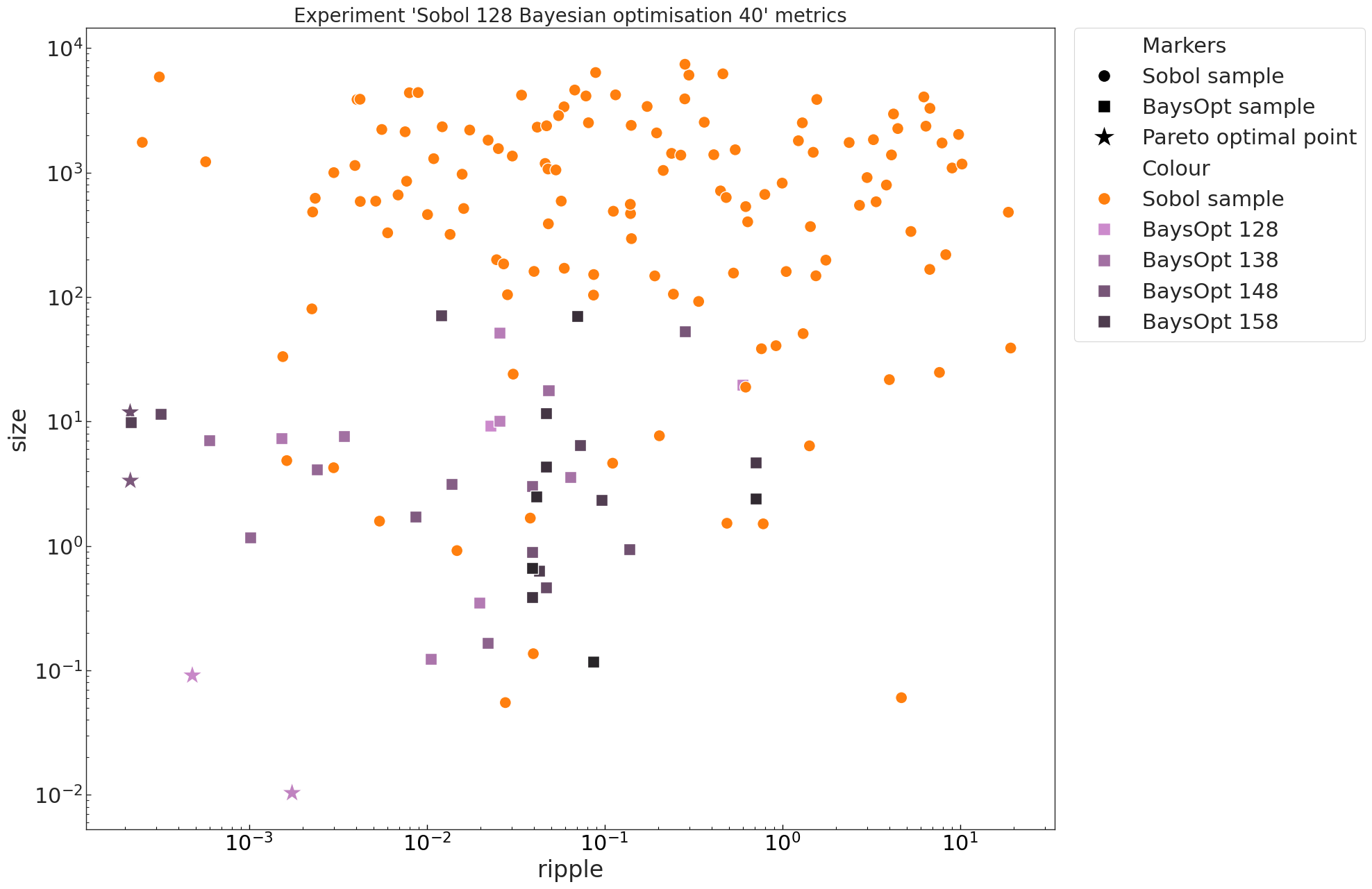}
    \caption{Log-log plot showing the resulting metrics of design points sampled in the Sobol stage (orange dots) and the Bayesian optimisation stage (purple squares) where the darker the colour of a Bayesian point, the later the iteration that design point corresponds to. Stars indicate the non-dominated points--the Pareto optimal design points.}
    \label{fig:both_metrics_and_trial_idx}
\end{figure}

Figure \ref{fig:both_metrics_and_trial_idx} shows the \emph{size} and \emph{ripple} values observed in experiment 2, it can be seen how a majority of the Sobol samples are poor quality design points. The Bayesian optimisation is immediately able to identify regional proximity of the the Pareto front, and descends towards it.

The results in Table \ref{tab:sobol_baysopt_comp} and Figure \ref{fig:both_metrics_and_trial_idx} are in line with the physical description of the problem as stated in section \ref{sec:problem}, and the identified design points are well within the expected range. By definition, a larger TF coil has a lower ripple, and vice versa. The parameterisations for experiment 2 and the corresponding designs are discussed further in Appendix \ref{app:appendix_resulting_designs}. 

\subsection{Comparison with other methods}

We benchmarked the Bayesian optimisation over a Gaussian Process surrogate approach against a purely quasi-random sampling across the search domain using Sobol, and a multi-objective evolutionary algorithm, specifically the Non-dominated Sorting Genetic Algorithm (NSGA-II) \cite{nsga2}. The NSGA-II is the de-facto optimisation toolkit used within Bluemira and is often the primary choice for determining design choices. We will compare each optimisation method with respect to a baseline experiment, experiment 2 in Table \ref{tab:sobol_baysopt_comp}. The comparison will be between the quality of the best Pareto optimal point(s) and the time taken to find said points for a fixed number of design point evaluations (168). All tests were run on a single core of an Intel(R) Xeon(R) Silver 4108 CPU @ 1.80GHz with 4GB of memory available.

\begin{table}[h!]
    \centering
    \begin{tabular}{|c|c|c|c|}
         \hline
         Method & size & ripple & time (s) \\
        \hline\hline
            Sobol + Bayesian optimisation (baseline) & 0.012 & 4.5e-4 & 872 \\
        \hline
            Sobol & 0.055, 0.43 & 0.028, 5.9e-3 & 691 \\
        \hline
            NSGA-II & 89.1 & 0.19 & 786 \\
        \hline
    \end{tabular}
    \caption{Comparison of the optimal point(s) for several multi-objective optimisation methods after allowing 168 \emph{ripple} function evaluations. Where multiple Pareto optimal points are deemed to be of the same quality, they are presented comma delimited.}
    \label{tab:method_comparison_fixed_number_evals}
\end{table}

The Sobol samples in Table \ref{tab:method_comparison_fixed_number_evals} are decent quality design points, but do not recover the Pareto frontier accurately. The best Sobol design points are still orders of magnitude larger than the best point in the baseline experiment. We can qualitatively verify this via Figure \ref{fig:both_metrics_and_trial_idx} where the Pareto frontier comprises only MOBO samples. Further, only a few of the Sobol samples are better than even the worst MOBO samples, with a majority of the Sobol samples orders of magnitude larger the MOBO samples. Overall, the best Sobol design points are of similar magnitude and quality to experiment 1 in Table \ref{tab:sobol_baysopt_comp}, which required half the number of function evaluations. 

We conclude that the 23\% increase in the run-time of the baseline MOBO with respect to the Sobol is acceptable, considering the significant improvement in design quality. The difference in time between Sobol and the baseline in Table \ref{tab:method_comparison_fixed_number_evals} can be attributed to the overhead of fitting the Gaussian Process and optimisation of the acquisition function. In order to achieve comparable design quality to the baseline experiment, more Sobol samples would need to be taken. Therefore, there exists a point where the the overhead of fitting the GP will be equal to, or less than, the time taken to evaluate the additional Sobol samples required to achieve the same design quality. This situation becomes more likely if the cost of evaluating the \emph{ripple} is increased by increasing the fidelity of the model and evaluating more ripple points on the plasma surface. We therefore find that reducing the number of function evaluations is significantly more important than the minor overhead of training. 

NSGA-II significantly under performs compared to both the baseline and Sobol. This is likely due to the curse of dimensionality hampering the evolutionary nature of NSGA-II. In order to efficiently explore the search space for the genetic algorithm, we would require a lot more iterations and evaluations before obtaining comparable values. The method used just 8 Monte Carlo points with 20 evaluations, rendering the optimiser unable to efficiently minimise across both the objectives. 

\section{Conclusion and Future Work} \label{conclusion}

Through this work, we demonstrate a proof-of-concept for a multi-objective Bayesian optimisation strategy deployed for determining better design strategies for a fusion reactor. Our work demonstrates that through MOBO we are able to better capture the Pareto front associated with the optimisation efficiently. The performance of the MOBO is further elucidated by benchmarking against a quasi-random exploration scheme (Sobol) and an evolutionary algorithm (NSGA-II). Though the optimised parameters lead to unrealistic design parameters for actual reactor construction, they accurately represent the ideal parameters as prescribed by Bluemira for the problem setup that we have provided, validating our optimisation strategy. As the next phase of our research, we would be extending the optimisation problem to include the magnetic equilibrium generated by the currents within the TF coils, optimising across both design and control elements of the reactor.


    
    
    


\newpage
\bibliographystyle{unsrt}
\bibliography{references}

\newpage
\appendix

\section{Tokamak Illustration}
\label{app:appendix_tokamak}
\begin{figure}[H]
    \centering
    \includegraphics[width=0.5\textwidth]{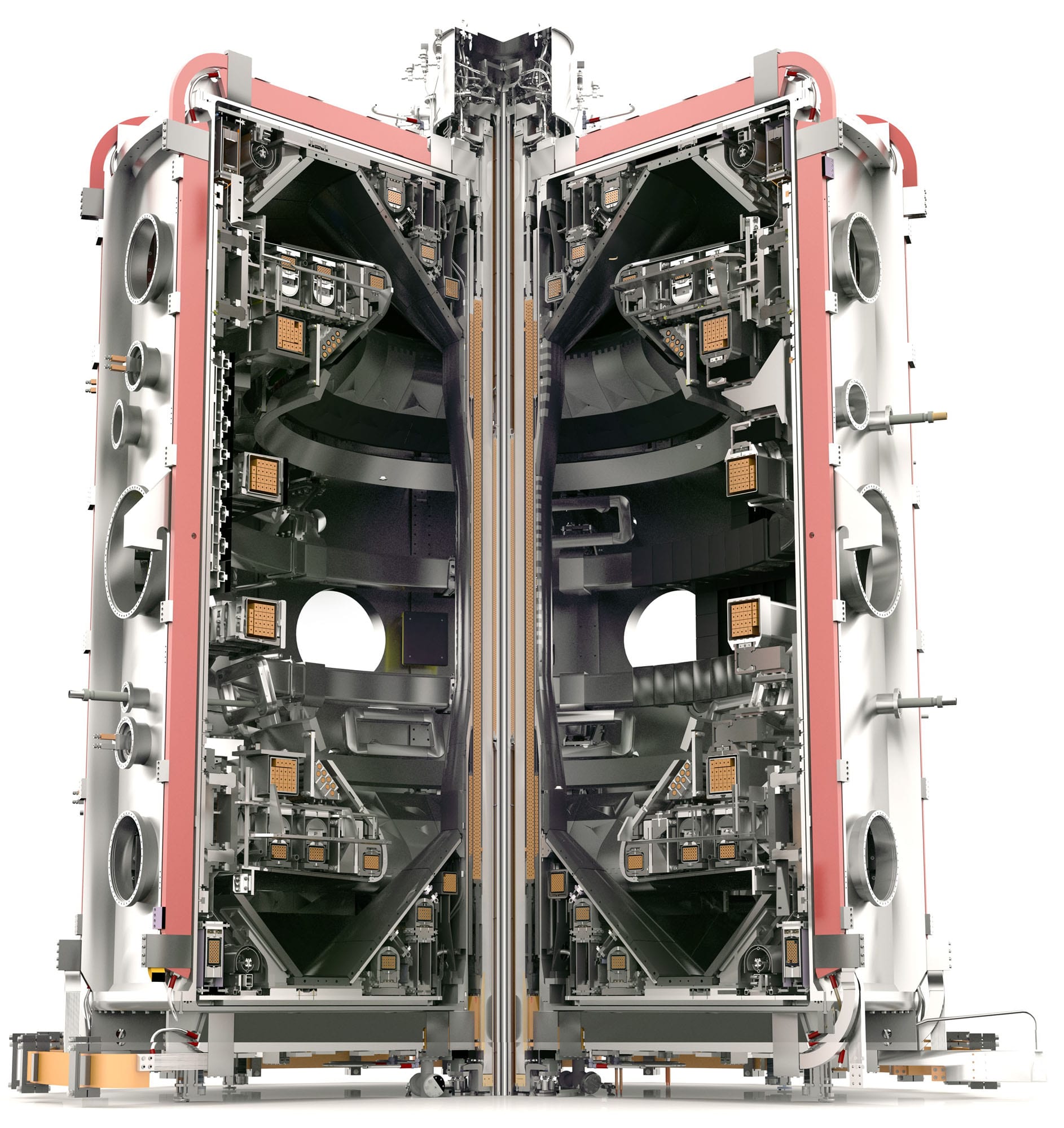}
    \caption{A cutaway illustration of MAST Upgrade, one of the Tokamaks at Culham Science Centre. The Toroidal Field coils (red) have a picture frame shape. Credit to UK Atomic Energy Authority.}
    \label{fig:app:JET_cutaway_1982}
\end{figure}

\begin{figure}[H]
    \centering
    \includegraphics[width=0.7\textwidth]{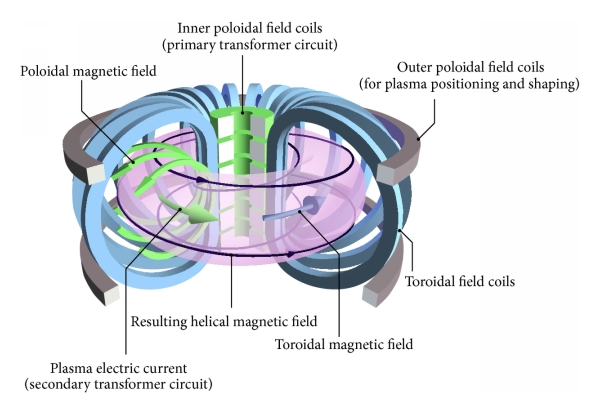}
    \caption{\cite{tokamak_schema} A simple schematic diagram of a generic Tokamak with all of the main magnetic components and fields shown. The TF coil magnets (blue) are that which this work aimed to optimise for.}
    \label{fig:app:JET_cutaway_1982}
\end{figure}

The vacuum vessel is the empty torus-shaped area at the center of the device where the plasma is magnetically confined. The toroidal magnetic field is produced by the Toroidal Field coils and is partially responsible for magnetic confinement of the plasma.

\newpage
\section{Parameter description and ranges}
\label{app:appendix_params}

\begin{figure}[H]
    \centering
    \includegraphics[width=0.35\textwidth]{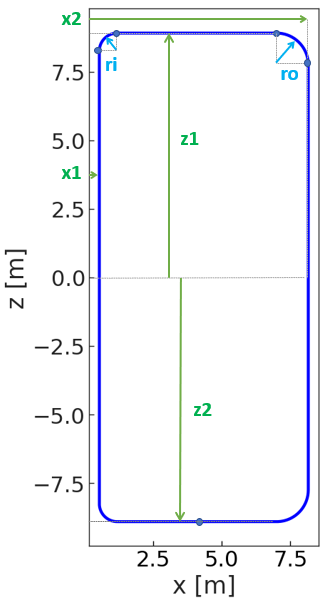}
    \caption{Illustration of the TF coil centreline shape parametrization.}
    \label{fig:app:TF_coil_shape_parametrization}
\end{figure}

\begin{figure}[H]
    \centering
    \includegraphics[width=0.90\textwidth]{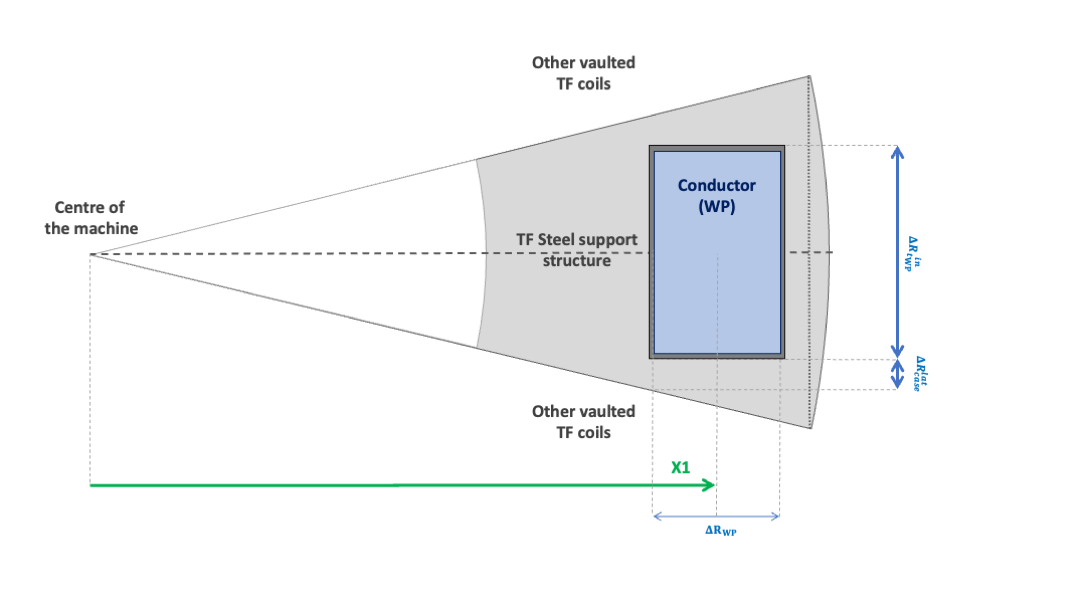}
    \caption{Illustration of the TF coil centreline shape parametrization.}
    \label{fig:app:TF_coil_xsection_parametrization}
\end{figure}

This appendix summarises the optimization variables and their ranges listed in Table~\ref{tab:var_description}. The parameters setting the TF coil dimensions, $\left\{x_1, x_2, z_1, z_2 \right\}$ are illustrated with the green arrows in figure~\ref{fig:app:TF_coil_shape_parametrization}, the corner rounding are parametrized by the light blue arrows $\left\{r_i, r_o \right\}$. The conductor cross-section is shown as the blue area in figure~\ref{fig:app:TF_coil_xsection_parametrization} showing the inboard mid-plane cross-section of a TF coil with a typical support structure (steel casing, grey). The rectangular cross-section is centred on the coil illustrated by the blue line in figure~\ref{fig:app:TF_coil_shape_parametrization}. Its radial width is taken as $\Delta R_\mathrm{WP}$, while its thickness in the toroidal direction $\Delta R_\mathrm{t, WP}^\mathrm{int}$ is parametrized with $f_{\Delta y_\mathrm{wp}}$ defined as 

$$  
f_{\Delta y_\mathrm{wp}} = \frac{\Delta R_\mathrm{t, WP}^\mathrm{int}}{\left(x_1 - \frac{\Delta R_\mathrm{t, WP}^\mathrm{int}}{2}\right) \tan\left(\frac{\pi}{N_\mathrm{TF}}\right)}
$$

to avoid overlapping conductors in the inboard regions\footnote{If $f_{\Delta y_\mathrm{wp}}$ is below 1, no overlap happens by definition.}. 

\begin{table}[h!]
\begin{center}
    \begin{tabular}{|c|c|c|}
    \hline
    Input & Description & range \\
    \hline\
        $N_\mathrm{TF}$ & Number of TF coils & $\left[16, 24 \right]$\\
    \hline
        $x_1$ & Inboard mid-plane TF coil centreline radius & $\left[0.0,  -4.5 \right]$ m \\
    \hline
        $x_2$ & Outboard mid-plane TF coil centreline radius & $\left[13.0, 20.0 \right]$ m \\
    \hline
        $z_1$ & Top TF coil centreline height & $\left[6.0, 15.0 \right]$ m\\
    \hline
        $z_2$ & Bottom TF coil centreline height & $\left[-15, -6.0 \right]$ m\\
    \hline
        $r_\mathrm{out}$ & Top/bottom outboard corners bending radii & $\left[0.0, 5.0 \right]$ m\\
    \hline
        $r_\mathrm{in}$ & Top/bottom inboard corners bending radii & $\left[10^{-5}, 0.14  \right]$ m\\
    \hline
        $\Delta R_\mathrm{WP}$ & Thickness of the conductor (winding pack) & $\left[0.1, 0.7  \right]$ m\\
        & in the poloidal plane & \\
    \hline
        $f_{\Delta y_\mathrm{wp}}$ & Thickness of the conductor (winding pack) & \\
        & in the toroidal direction as a fraction of the & $\left[0.1, 1.0 \right]$ \\  
        & maximum allowable value & \\
    \hline
    \end{tabular}
    \caption{Description of the input parameters for the optimization.}
    \label{tab:var_description}
\end{center}
\end{table}

\newpage
\section{Additional Pareto frontiers}
\label{app:addition_pareto_front}
This appendix includes various Pareto frontier plots for experiments in Tables \ref{tab:sobol_baysopt_comp} \& \ref{tab:method_comparison_fixed_number_evals}.
\begin{figure}[H]
    \centering
    \includegraphics[width=0.7\textwidth]{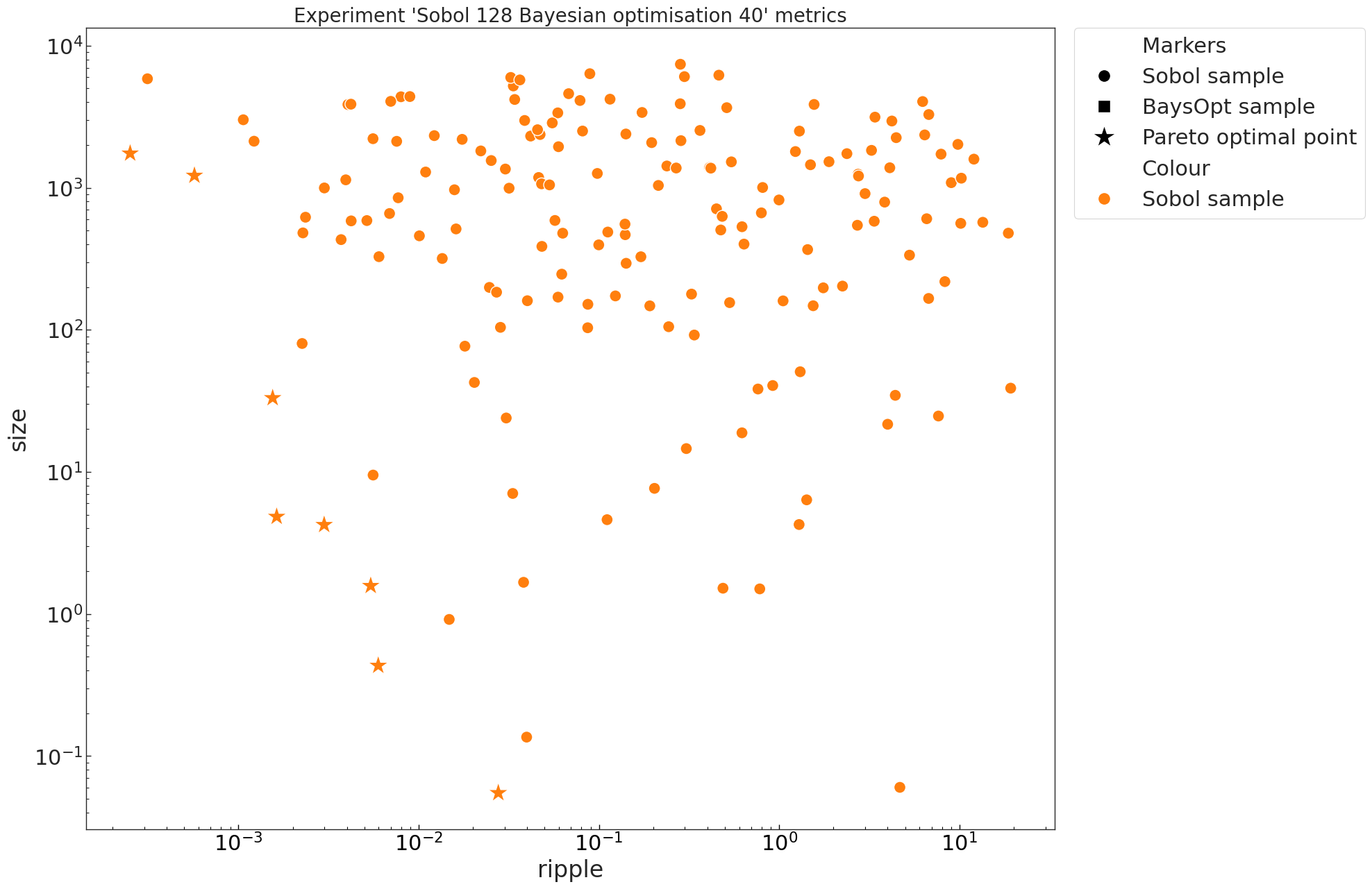}
    \caption{shows the Pareto frontier of the Sobol method from Table \ref{tab:method_comparison_fixed_number_evals}. When compared with Figure \ref{fig:both_metrics_and_trial_idx} it can be seen that the Pareto frontier exists several orders of magnitude above where it does for the baseline in Table \ref{tab:method_comparison_fixed_number_evals}}
    \label{fig:sobol_frontier}
\end{figure}

\begin{figure}[H]
    \centering
    \includegraphics[width=0.7\textwidth]{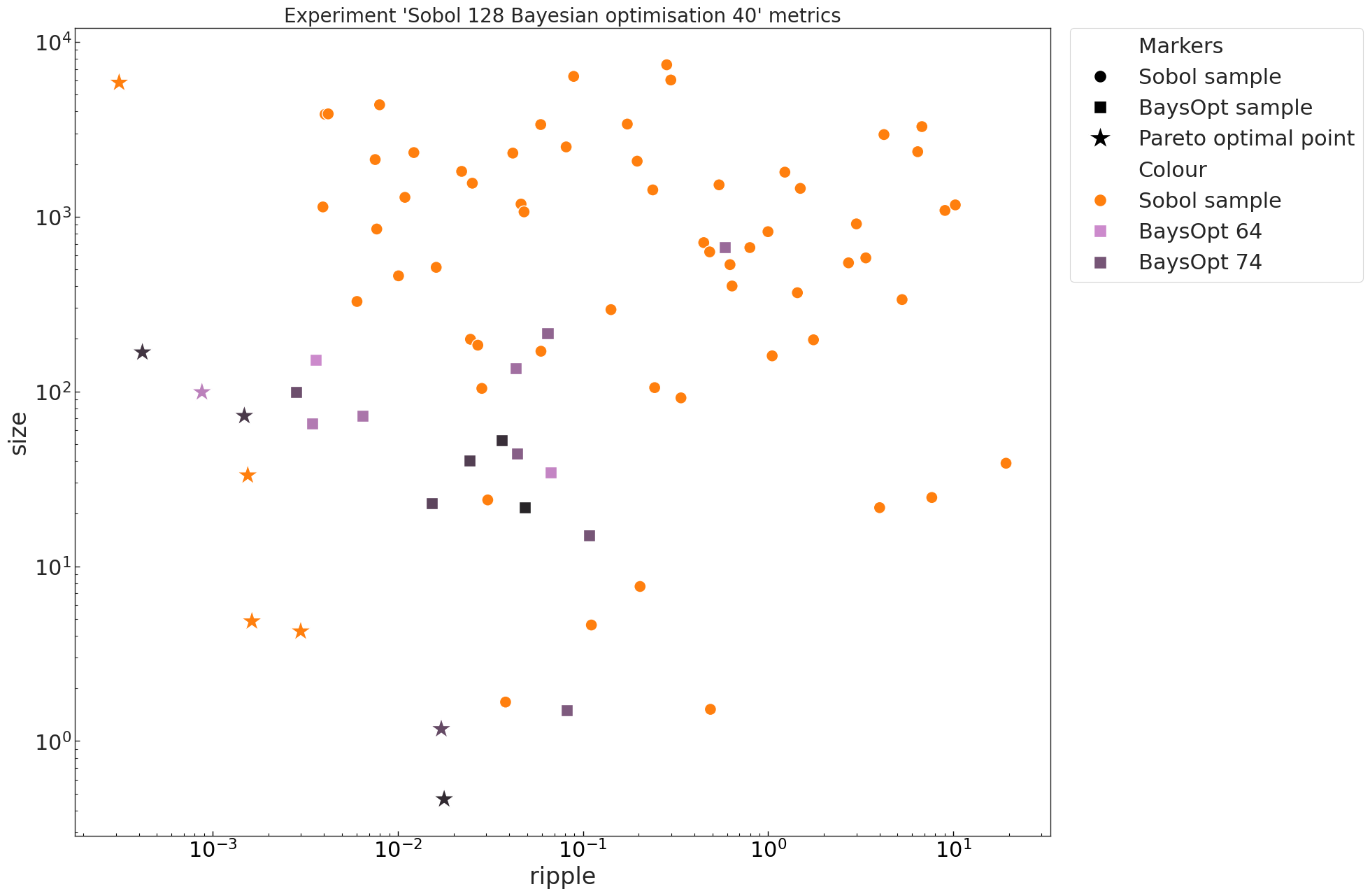}
    \caption{shows the Pareto frontier and distribution of metrics for experiment 1 of Table \ref{tab:sobol_baysopt_comp}. We can see that the frontier is similar to that shown in \ref{fig:sobol_frontier} despite this frontier needing half the number of function evaluations to reach.}
    \label{fig:smaller_sobol_baysopt_frontier}
\end{figure}

\newpage
\section{Experiment 'Sobol 128 + Bayesian optimisation 40' optimal design point}
\label{app:appendix_resulting_designs}

Below we briefly discuss the design point deemed to be of the best quality of the four Pareto optimal design points in experiment 2 in Table \ref{tab:sobol_baysopt_comp}, sample 129 of 168. We include the parameter values of the design point and include a picture of the resulting coil.

\begin{table}[H]
    \centering
    \begin{tabular}{|c|c|}
        \hline
        Parameter & Value \\
        \hline\hline
        dr wp & 0.1m \\
        f y wp & 0.659 \\
        number of coils & 25 \\
        x1 & 0.0348m \\
        x2 & 19.35m \\
        z & 15.0m \\
        ri & 0.0931m \\
        ro & 5.0m \\
        \hline
    \end{tabular}
    \caption{parameterisation of design point 19 for experiment 2}
    \label{tab:129_params}
\end{table}

\begin{figure}[H]
    \centering
    \includegraphics[width=0.65\textwidth]{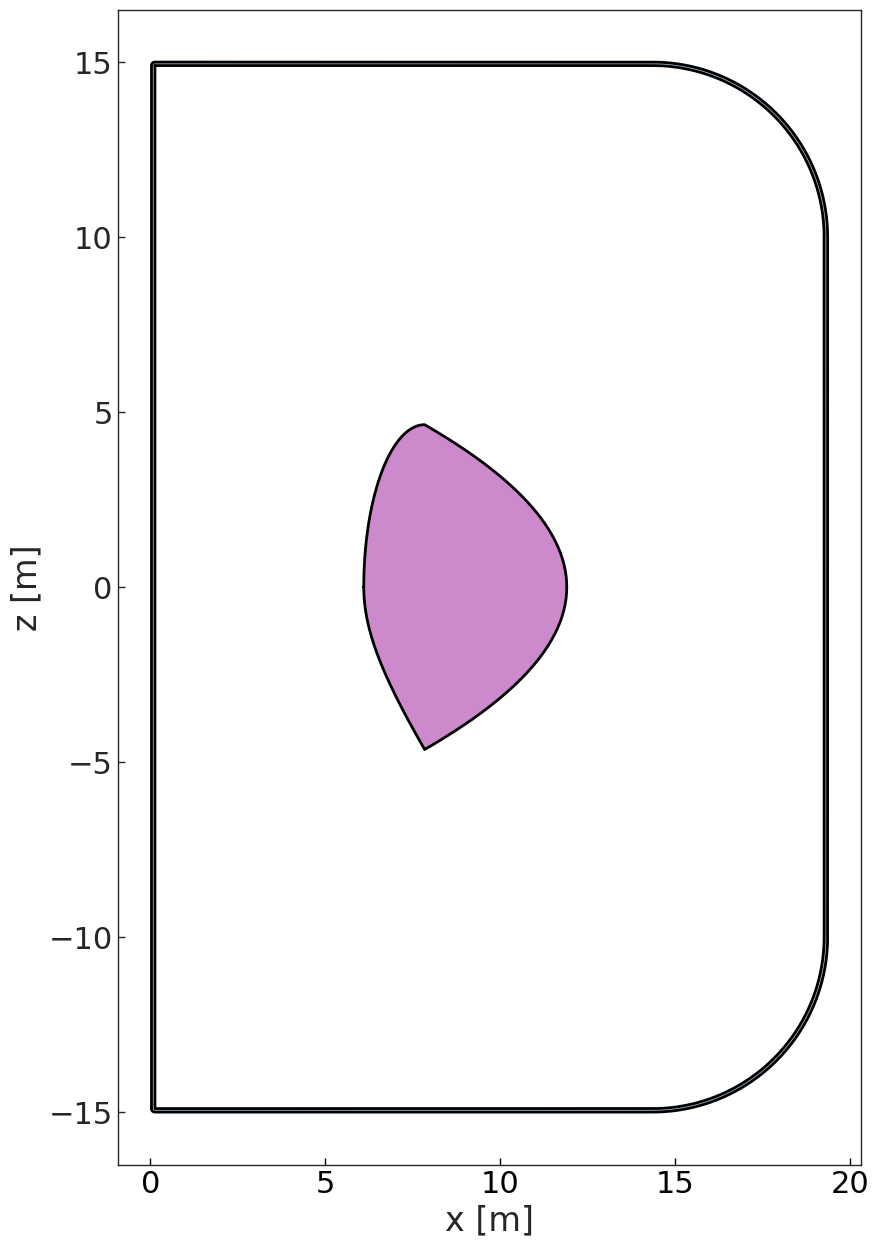}
    \caption{resulting TF coil (blue) from the parameterisation in Table \ref{tab:129_params} with the fixed plasma shape (purple) shown for reference.}
    \label{fig:129_figure}
\end{figure}

This coil is rather large, ensuring that the distance $x2 - x1$ is maximised as this reduces the \emph{ripple} the most. This means $x1$ is close to its lower bound, and $x2$ is close to its upper bound. The algorithm manages to achieve a small volumetric size of the TF coil by making the winding pack extremely thin to the point where this coil would likely be physically unstable. Future work, discussed in Section \ref{conclusion}, will aim to resolve this by enforcing constraints on stresses experienced by the coil structure.

\section{Experiment 'Sobol 128 + Bayesian optimisation 40' Gaussian Process Fits}
\label{appendix: gp_fit}
\begin{figure}[H]
    \centering
    \includegraphics[width=1\textwidth]{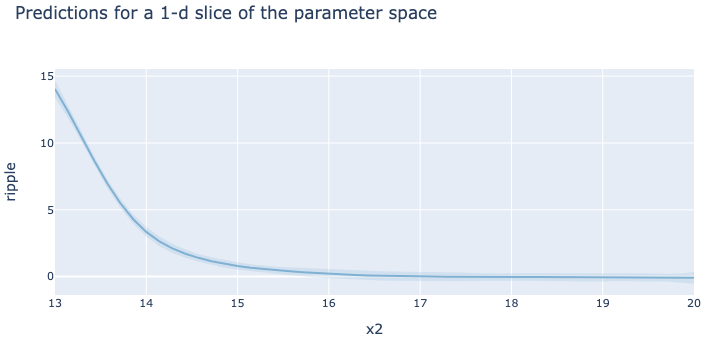}
    \caption{shows the fit of the Gaussian Process for parameter $x2$ on the \emph{ripple} metric given that the other variables have been fixed at their midpoints.}
    \label{fig:x1_ripple_slice}
\end{figure}

\begin{figure}[H]
    \centering
    \includegraphics[width=1\textwidth]{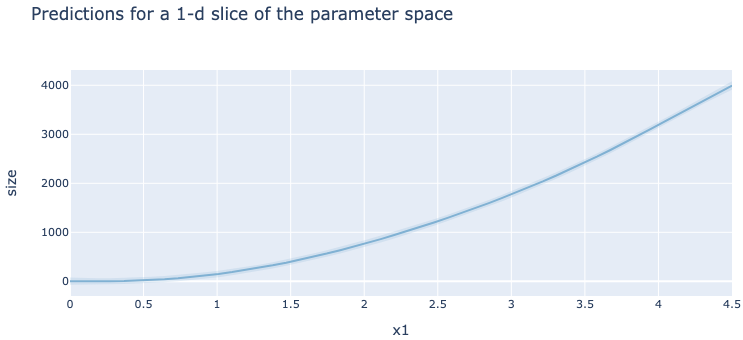}
    \caption{shows the fit of the Gaussian Process for parameter $x1$ on the \emph{size} metric given that the other variables have been fixed at their midpoints.}
    \label{fig:x1_size_slice}
\end{figure}

Figures \ref{fig:x1_ripple_slice} \& \ref{fig:x1_size_slice} shows two of the Gaussian Process fits for $x2$ on the \emph{ripple} metric and for $x1$ on the \emph{size} metric, respectively. We can see that the Gaussian Process' are extremely confident in both metrics given these two physically relevant parameters.

\end{document}